\documentclass[9pt,twocolumn,twoside]{opticajnl}
\journal{opticajournal} 

\setboolean{shortarticle}{true}

\title{Resolved Raman sideband cooling of a single optically trapped cesium atom }

\author[1,2]{Zhuangzhuang Tian}
\author[1,2]{Haobo Chang}
\author[1,2]{Xin Lv}
\author[1,2]{Mengna Yang}
\author[1,2]{Zhihui Wang}
\author[1,2]{Pengfei Yang}
\author[1,2]{Pengfei Zhang}
\author[1,2,*]{Gang Li}
\author[1,2,$\dagger$]{Tiancai Zhang}

\affil[1]{State Key Laboratory of Quantum Optics and Quantum Optics Devices, Institute of Opto-Electronics, Shanxi University, Taiyuan 030006, China}
\affil[2]{Collaborative Innovation Center of Extreme Optics, Shanxi University, Taiyuan, Shanxi 030006, China}

\affil[*]{gangli@sxu.edu.cn}
\affil[$\dagger$]{tczhang@sxu.edu.cn}

\begin{abstract}
We developed a resolved Raman sideband cooling scheme that can efficiently prepare a single optically trapped cesium (Cs) atom in its motional ground states. A two-photon Raman process between two outermost Zeeman sublevels in a single hyperfine state is applied to reduce the phonon number. Our scheme is less sensitive to the variation in the magnetic field than the commonly used scheme where the two outermost Zeeman sublevels belonging to the two separate ground hyperfine states are taken. Fast optical pumping with less spontaneous emission guarantees the efficiency of the cooling process.
After cooling for 50 ms, 82\% of the Cs atoms populate their three-dimensional ground states. Our scheme improves the long-term stability of Raman sideband cooling in the presence of magnetic field drift and is thus suitable for cooling other trapped atoms or ions with abundant magnetic sublevels.
\end{abstract}

\setboolean{displaycopyright}{false} 

\begin{document}

\maketitle

Optically trapped single atoms are one of the key building blocks for quantum information processors \cite{Saffman_algorithms,Lukin_algorithms}, quantum simulators \cite{2D_antiferromagnets,spin_liquids,continuous_symmetry_breaking}, quantum memory devices \cite{Rempe_atom_memory,Rempe_photon_memory}, and quantum metrology devices \cite{Kaufman_3s_clock,Kaufman_30s_clock,Kauffman_spin_squeeze,Antoine_spin_squeeze}. Advances in quantum manipulation precision in these applications set stringent requirements for full control of single atoms. Due to the coupling between the internal quantum states and the motional states in optical traps \cite{Diter_dephasing}, the precise control of the atom motional states becomes one of the key factors for improving the precision of quantum manipulations. Preparation of the motional ground states, also termed the zero phonon states, is usually the first step in controlling the motional states \cite{Wineland_Nobel_RMP}. 

One of the common ways to obtain the zero phonon state is to cool the optically trapped atom by Raman sideband cooling (RSC) \cite{Anderson_RSC,HeXD_RSC,Kaufman_RSC,Lukin_RSC,Wang_21,Weiss_projection_cooling,Spence_2022,Diter_statedependent_lattice,PhysRevLett.129.103401} due to its high efficiency compared to other sub-Doppler cooling methods.
There are usually two types of RSC, i.e., the resolved RSC \cite{Anderson_RSC,HeXD_RSC,Kaufman_RSC,Lukin_RSC} and projection RSC \cite{Weiss_projection_cooling,Diter_statedependent_lattice}, both of which are carried out in two steps. The first step is the transfer of the phonon state from $|n\rangle$ to $|n-1\rangle$, which is achieved with the aid of an electronic state transition. The second step is optical pumping, in which the atom is pumped back to its initial electronic state without changing the phonon state, this requires that the effective Lamb-Dicke parameter be small enough. 
In the resolved RSC, the first step is realized by a two-photon Raman process that drives an electronic transition with $\Delta n = -1$ sideband. A high transfer efficiency requires resolvable $\Delta n = \pm 1$ sidebands from the $\Delta n = 0$ carrier.
In the projection RSC \cite{Diter_statedependent_lattice,Weiss_projection_cooling}, the two sets of phonon wavefunctions associated with two electronic states are deliberately offset in space. Thus, the first step can be realized by a microwave driving field. 

The fidelity of the obtained zero phonon state strongly depends on the efficiency of optical pumping \cite{Wineland_RSC_RMP}. The less spontaneous scattering, the higher the fidelity for a limited Lamb-Dicke parameter. Most of the experiments adopt two outermost Zeeman sublevels belonging to two separate ground hyperfine states, e.g., $|F=3, m_F=\pm 3 \rangle$ and $|F=2, m_F=\pm2 \rangle$ for rubidium-85 \cite{Kaufman_RSC,Lukin_RSC,HeXD_RSC}, $|F=2, m_F=\pm 2 \rangle$ and $|F=1, m_F=\pm1 \rangle$ for rubidium-87 \cite{HeXD_RSC}, and $|F=4, m_F=\pm 4 \rangle$ and $|F=3, m_F=\pm3 \rangle$ for cesium (Cs) \cite{Diter_statedependent_lattice,Weiss_projection_cooling,Spence_2022}. However, the atomic transitions between the two electronic states are susceptible to the magnetic field. The drift of the magnetic field will offset the atomic transition from the $\Delta n=-1$ driving field and ultimately degenerate the RSC efficiency. To suppress this effect, electronic states with $m_F=0$ have been proposed for executing the RSC \cite{Anderson_RSC}, but the spontaneous scattering in the optical pumping process increases substantially.

In this letter, we propose an improved scheme of resolved RSC in which the two outermost Zeeman sublevels in a single hyperfine state, i.e., $|F=4, m_F= 4 \rangle$ and $|F=4, m_F=3 \rangle$ of Cs in our experiment, are chosen.
The scheme greatly mitigates the influence of magnetic field drifts and avoids increasing spontaneous scattering during optical pumping. 
We use this scheme to prepare the zero-phonon state of an optically trapped Cs atom with high efficiency. We finally obtain a population of 0.82 for the single Cs atom in its three-dimensional (3D) vibrational ground states. 
Our work paves the way for the precise manipulation of single Cs atoms or single Cs atom arrays for quantum science investigations.

\begin{figure}[t]
\centering
\includegraphics[width=\columnwidth]{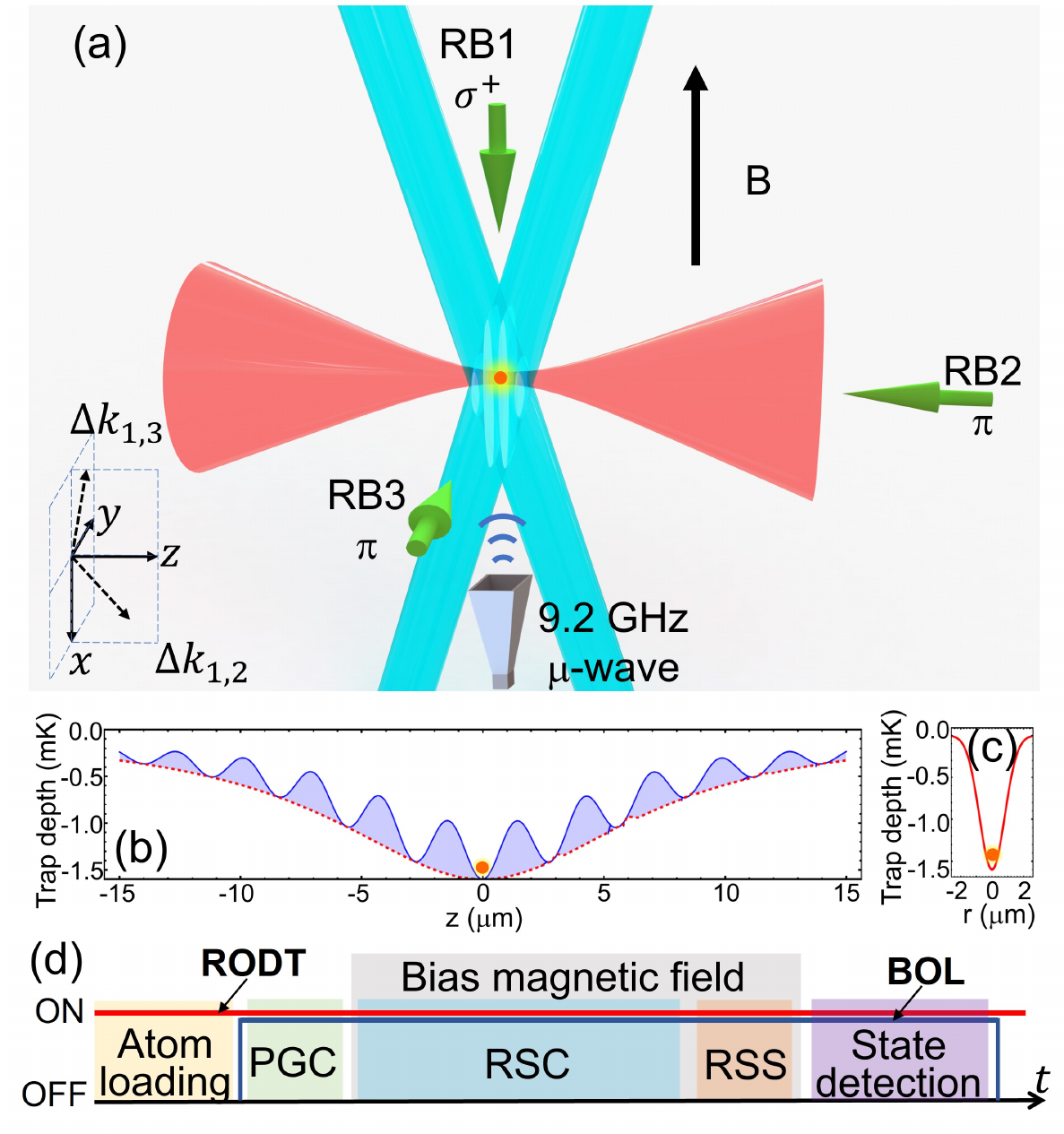}
\caption{(a) Sketch of the experimental setup. The three-dimensional tight optical trap for the RSC of a single Cs atom is obtained by combining a red-detuned 1052-nm trap (red beams, RODT), which provides tight confinement on the $x$-- and $y$--axes, and a blue-detuned 780-nm optical lattice (blue beams, BOL), which provides tight confinement on the $z-$axis. The BOL is formed by crossing two light beams at small angles. Three Raman beams (RB1, RB2, and RB3) are used to realize 3D cooling of the atoms. A magnetic field B along the $-x$ direction is used to define the quantization axis. A 9.2 GHz microwave field is used to apply a $\pi$-pulse between ground states $|F=4, m_F=4\rangle$ and $|F=3, m_F=3\rangle$ [see Fig. \ref{scheme}(a)]. (b) and (c) show the profile of the combined optical trap along the axial ($z$) direction and the radial ($x$ and $y$) direction, respectively. (d) shows the time sequence for one iteration of the experiment. RSS: Raman sideband spectroscopy.}
\label{setup}
\end{figure}

Figure \ref{setup}(a) shows a sketch of our experimental setup. A red-detuned optical dipole trap (RODT) is used to load single atoms from a cold atom ensemble prepared by a magneto-optical trap (MOT). A trap with waists of $(w_x,w_y)=(1.7,1.6)$ $\mu$m is obtained by strongly focusing a 1052-nm beam through an objective lens with $\text{NA}=0.4$. By using 35 mW of light, an optical trap with a depth of $-1.6$ $mK$ is formed. The trap provides tight confinement of a single Cs atom along the radial directions [Fig. \ref{setup}(c)], and the trap frequencies are $(\omega_x,\omega_y)=2\pi \times (59.5,69.7)$ kHz. However, the confinement on the z-axis is much looser. The trap frequency is $\omega_z=2\pi \times 7.2$ kHz, which is not suitable for the RSC. To perform the RSC, we utilize a blue-detuned optical lattice (BOL) to increase the confinement in the axial direction. 
The lattice is formed by crossing a pair of 780-nm light beams with an angle of $18^{\circ}$, which gives a lattice constant $a=2.7$ $\mu$m. The waist of the 780-nm beam is 18 $\mu$m. As shown in Fig. \ref{setup}(a), the two beams are aligned to cross at the position of the waist and overlap with the RODT.
The potential of the combined trap along the axial direction is shown in Fig. \ref{setup}(b). 
Thus, the trap frequency along the axial direction can be increased to $\omega_z = 2 \pi \times 32.3$ kHz with a power of 240 mW for each 780-nm laser beam. Thus far, a 3D tight trap has been constructed. 

The time sequence for the experiment is shown in Fig. \ref{setup}(d). A single Cs atom is first loaded by the RODT alone, and then the BOL is switched on. Therefore, only one atom is captured in the combined trap, and the atom is located at the center of the RODT [Fig. \ref{setup}(b) and (c)], which guarantees the same trap frequencies for every run of the experiment.
A subsequent 5-ms polarization gradient cooling phase is applied to cool the atom to a temperature of $\sim$ 10 $\mu$K. Then, an RSC pulse sequence is applied, and Raman sideband spectroscopy is performed to evaluate the residual phonon numbers.

\begin{figure}[t]
\centering
\includegraphics[width=\columnwidth]{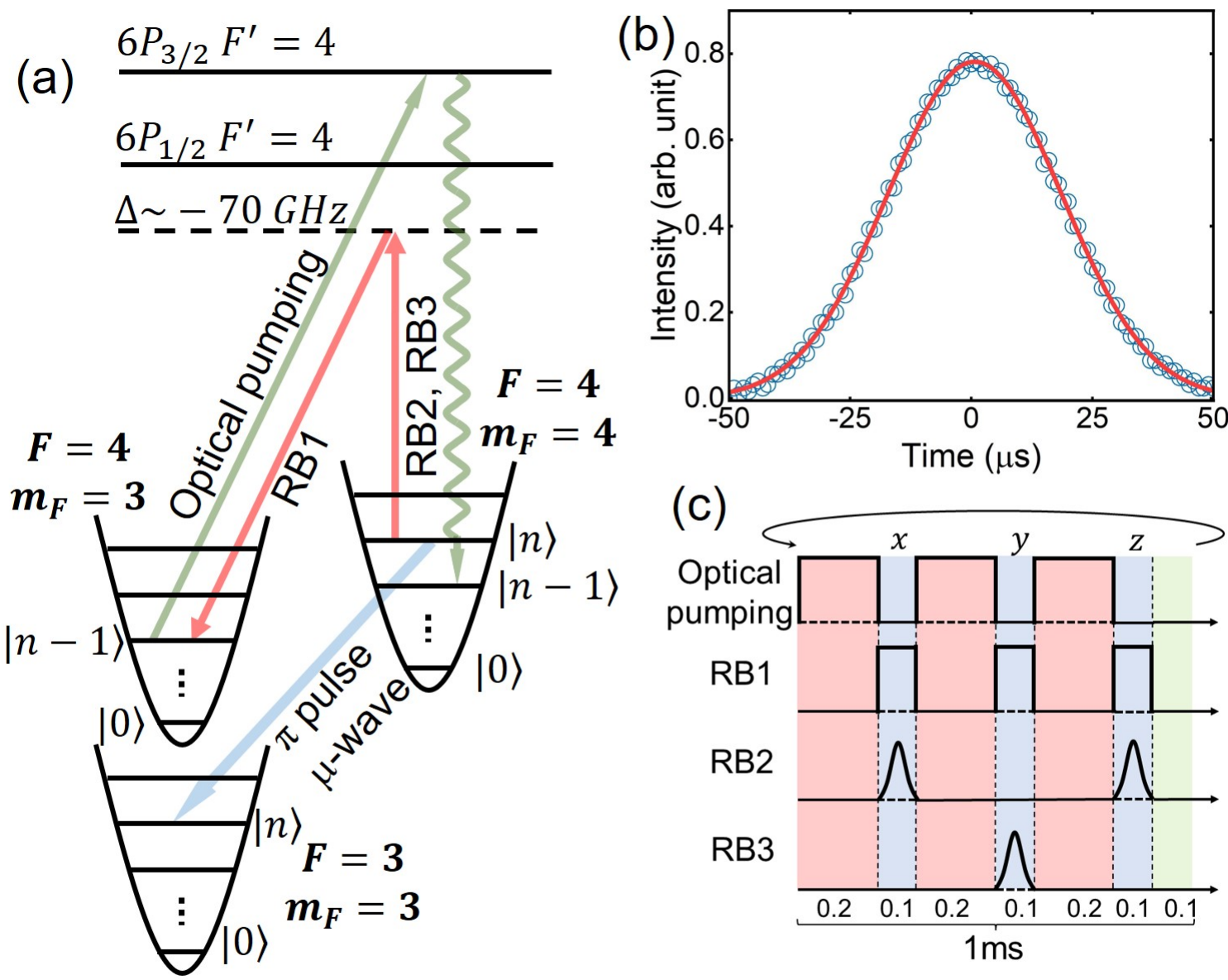}
\caption{(a) Energy-level diagram for the RSC. the Raman lasers are $-70$ GHz below the hyperfine level $6P_{1/2}, F^{\prime}=4$. The atomic transition between $|4,4\rangle$ and $|4,3\rangle$ is adopted to execute the RSC. 
An optical pumping light on-resonant with $|6S_{1/2},F=4\rangle \rightarrow |6P_{3/2},F=4^{\prime}\rangle$ transition in addition to repumping light is used to pump the atom from $|6S_{1/2}, F=4,m_F=3\rangle$ back to $|6S_{1/2}, F=4,m_F=4\rangle$. To finish the state detection, the atoms in $|6S_{1/2},F=4, m_F=4\rangle$ were transferred to $|6S_{1/2}, F=3, m_F=3\rangle$ with the aid of a microwave $\pi$-pulse. (b) A typical Gaussian profile for RB2 and RB3. The black dots are the experimental data, and the red curve is fitted by a Gaussian function $I(t)=I^0 \exp{[-t^2/(2\tau^2)]}$. The fitted $\tau= 18.3 \pm 0.7$ $\mu$s. (c) A cycle of cooling Raman pulses, in which RB1 remains in square form. The length of the Raman pulses was set to 100 $\mu$s. Three optical pumping pulses with a length of 200-$\mu$s are inserted before every pair of Raman pulses. The whole cycle lasts for 1 ms, and the final 100-$\mu$s time window is left blank.}
\label{scheme}
\end{figure}

The scheme of the RSC is shown in Fig. \ref{scheme}(a), where two Zeeman sublevels $|6S_{1/2}, F=4,m_F=4\rangle$ ($|4,4\rangle$) and $|6S_{1/2},F=4,m_F=3\rangle$ ($|4,3\rangle $) are adopted to execute the Raman process.
The energy difference between the two states is $\Delta E=\mu_{\mathrm{B}}g_{F}B$, where $\mu_{\mathrm{B}}$ and $g_{F}$ represent the Bohr magneton and Lande g-factor, respectively.
The dependence of $\Delta E$ on B is 1/7 of the RSC scheme in previous works \cite{Diter_statedependent_lattice,Weiss_projection_cooling,Spence_2022}, where $|6S_{1/2}, F=4,m_F=4\rangle$ and $|6S_{1/2},F=3,m_F=3\rangle$ are used and $\Delta E'={\hbar}\omega_{hfs}+7\mu_{\mathrm{B}}g_{F}B$ with $\omega_{hfs}$ the hyperfine splitting.
The two Zeeman states are driven by three Raman lasers (RB1, RB2 and RB3).
RB1 is $\sigma^{+}$-polarized and propagates along the $x$-axis. Because the B-field (quantization axis) is in the $-x$ direction, RB1 couples the atomic transition $|4,3\rangle \leftrightarrow |6P_{1/2}, F=4,m_F=4\rangle$. RB2 and RB3 propagate along the $-z$ and $y$ directions, respectively, with $\pi$ polarization. The two light beams couple the atomic transition $|4,4\rangle \leftrightarrow |6P_{1/2}, F=4,m_F=4\rangle$. The one-photon frequency detuning for all Raman lights is $\Delta=-70$ GHz to the corresponding atomic transitions. Therefore, the population on state $|6P_{1/2}, F=4,m_F=4\rangle$ is eliminated during the two-photon Raman process. 
The combination of RB1 and RB2 (RB1 and RB3) is used to cool the motion of atoms in the x and z directions (y direction).
The Lamb-Dicke parameter for Raman transition is defined as $\eta_q^{\mathrm{R}}\equiv q_{0}\Delta k_q$,
in which $q_0=(\hbar/2m\omega_q)^{1/2}$ ($q=x$, $y$ or $z$) is the oscillator length with $m$ the mass of the atom
and $\Delta k_q$ the momentum difference of two Raman beams on axis $q$.
In our experiment, $(\eta^{\mathrm{R}}_x, \eta^{\mathrm{R}}_{y}, \eta^{\mathrm{R}}_z)$ = $(0.16, 0.25, 0.23)$.

The RSC process begins with the atom initially in state $|4,4\rangle$ and phonon state $|n_q\rangle$. 
The trapped atom is initialized to $|4,4\rangle$ by optical pumping, which is achieved by a combination of two $\sigma^+$-polarized beams with frequencies resonating to atomic transitions $|6S_{1/2}, F=4\rangle \leftrightarrow |6P_{3/2}, F'=4\rangle$ and $|6S_{1/2}, F=3\rangle \leftrightarrow |6P_{3/2}, F'=4\rangle$). An additional optical pumping beam is also applied to pump atoms dropped at $|4,3\rangle $ back to $|4,4\rangle $ during the RSC. This process only takes an average of 2.1 spontaneous emissions.
The Lamb-Dicke parameters for the optical pumping process in the three directions are $(\eta_{x}^{OP},\eta_{y}^{OP},\eta_{z}^{OP})=(0.172, 0.186, 0.253)$, which guarantees that the phonon state is not altered during the optical pumping process.

After the state initialization, a Raman pulse sequence is applied on the atom to execute the RSC. A typical cycle of the pulse sequence can be found in Fig. \ref{scheme}(c). Every cycle takes one ms and includes three Raman pulses, which are applied sequentially to cool the atom motion on the $z$-, $y$-, and $x$-axes. The duration of each Raman pulse is 100 $\mu$s, and a 200-$\mu$s optical pumping pulse is added in advance. The last 100-$\mu$s time window is deliberately left blank. The cooling cycle repeats until the ground phonon states are reached.

For every single Raman pulse, it is set to be two-photon-resonant with transition $|4,4\rangle \otimes |n_q\rangle \rightarrow |4,3\rangle \otimes |n_q- 1\rangle$. The pulse drives the atom to state $|4,3\rangle \otimes |n_q- 1\rangle$, where the phonon number decreases by one. Then, optical pumping is applied to pump the atom back to state $|4,4\rangle$ with the same phonon state $ |n_q- 1\rangle$. This process repeats until the atom accumulates in the $ |4,4\rangle \otimes |0_q\rangle$ state and the RSC finishes. 
To avoid the off-resonantly driving of the transition $|4,4\rangle \otimes |n_q\rangle \rightarrow |4,3\rangle \otimes |n_q\rangle$ and 
to increase the cooling efficiency, a Gaussian Raman pulse \ref{scheme}(b), instead of the square wave pulse, is used to drive the $|4,4\rangle \otimes |n_q\rangle \rightarrow |4,3\rangle \otimes |n_q-1\rangle$ transition.
The reason is that the spectral shape of the Raman process obtained by a temporal Gaussian-shaped driving pulse still takes a Gaussian form, which is clearer than the sinc-function-shaped spectrum obtained by a temporal square-shaped driving pulse.

We shape RB2 and RB3 into a temporal Gaussian profile with intensity following the function $I_{2(3)}(t)=I^0_{2(3)}\exp{[-t^2/(2\tau^2)]}$, where $I^0_{2(3)}$ is the peak intensity and $\tau$ is the time waist. Therefore, the Rabi frequency of RB2 (RB3) takes the form $\Omega_{2(3)}(t)=\Omega^0_{2(3)}\exp{[-t^2/(4\tau^2)]}$ with $\Omega^0_{2(3)}$ being the peak Rabi frequency. RB1 uses a square wave pulse with Rabi frequency $\Omega_{1}$. The ``carrier'' (couples $|4,4\rangle \otimes |n_q\rangle \rightarrow |4,3\rangle \otimes |n_q\rangle$) Rabi frequency of the composed two-photon Raman process still takes the temporal profile $
\Omega_{c}(t)=\Omega^0_c \exp{[-t^2/(4\tau^2)]}$, with $\Omega^0_c=\frac{\Omega_1\Omega^0_{2(3)}}{2\Delta}$ being the maximum two-photon Rabi frequency.
The Rabi frequency of the two-photon Raman process for $|4,4\rangle \otimes |n_q\rangle \rightarrow |4,3\rangle \otimes |n_q- 1\rangle$ is
\begin{equation}
\Omega_{n_q-1}(t)\simeq\Omega_c(t)\eta_q^{\mathrm{R}}\sqrt{n_q},
\label{n=-1}
\end{equation}
where  $\eta_q^{\mathrm{R}}$ is the Lamb-Dicke parameter for Raman transition. 
$\Omega_{n_q-1}(t)$ also takes the Gaussian profile and is dependent on the phonon number. To resolve the $\Delta n =\pm 1$ band of the two-photon Raman transition we choose a constant Raman pulse length of $\tau = 25$ $\mu$s. The full duration of the Raman pulse is 100 $\mu$s. Figure \ref{scheme}(b) displays a typical Gaussian profile for RB2 and RB3.
We set the peak Rabi frequencies of the three Raman beams as $(\Omega_{1}, \Omega_{2}, \Omega_{3}$) = $2\pi\times (122.6, 28.6, 26.2)$ MHz; thus, the Gaussian-shaped Raman pulse works as a $\pi$-pulse for transition $|4,4\rangle \otimes |1_q\rangle \rightarrow |4,3\rangle \otimes |0_q\rangle$. The setting remains the same for the whole RSC process.

\begin{figure}[ht]
\centering
\includegraphics[width=\columnwidth]{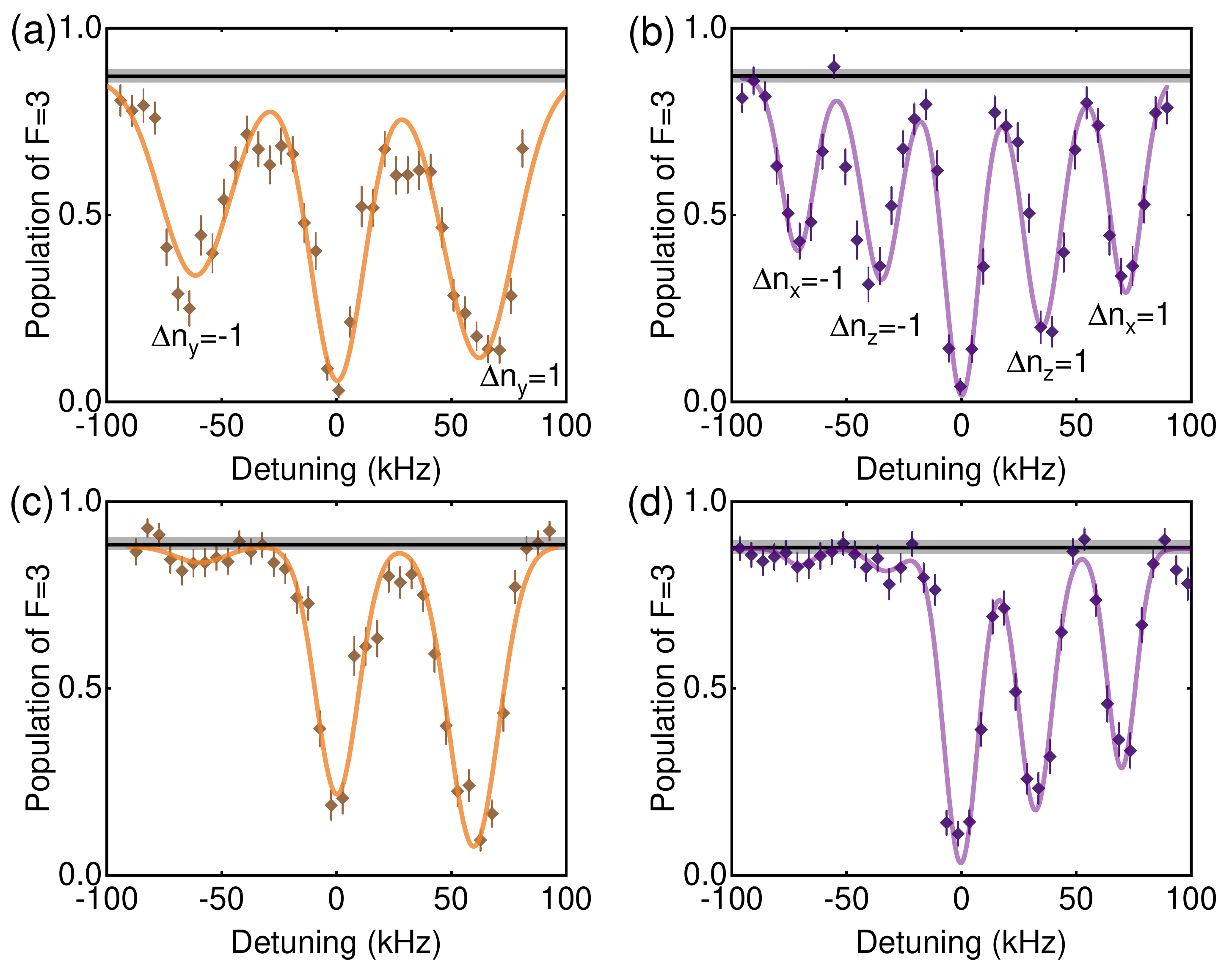}
\caption{Raman sideband spectra of the optically trapped single Cs atoms before [(a) and (b)] and after [(c) and (d)] the RSC, where the Raman cooling pulse is applied for 50 cycles. (a) and (c) display the spectra for the atomic motion on the $y$-axis. (b) and (d) show the spectra on the $x$- and $z$-axes.}
\label{spectroscopy}
\end{figure}

To evaluate the cooling efficiency, the Raman sideband spectra were measured before and after the RSC sequences.
The spectrum is obtained by measuring the population on $|4,4\rangle$ after applying a single Raman pulse (RB1 and RB2, or RB1 and RB3) with the frequency of RB1 scanned in a certain range. 
The optical pumping beam remains off during this process.
The peak Rabi frequency and the profile of the Raman pulse are the same as those of the RSC.
To measure the population of atoms in states $|4,4\rangle$, a resonant microwave $\pi$-pulse is applied between states $|4,4\rangle$ and $|6S_{1/2},F=3,m_F=3\rangle$. It transfers the atom from $|4,4\rangle$ to $|6S_{1/2},F=3,m_F=3\rangle$. The atom remaining in $|6S_{1/2}, F=4\rangle$ state will be blown away from the trap by a resonant light beam, while the atom in $|6S_{1/2}, F=3,m_F=3\rangle$ state stays. Then, the MOT beams are switched on, and the photons scattered from the atom are recorded by a single photon counter. The population on $|4,3\rangle$ is obtained by statistics of the atom stay events for approximately 100 trials of the experiment. The sideband spectra of the atoms on the $x$- and $z$-axes are obtained by applying RB1 and RB2, and the sideband spectrum on the $y$-axis is obtained by applying RB1 and RB3.

The sideband spectra before the RSC are measured first to evaluate the initial average photon number on the three axes. The results are shown in Fig. \ref{spectroscopy}(a) and (b).
The large peaks on the $\Delta n_q=-1$ sideband indicate the large phonon numbers on the three axes.
The average phonon number $\bar{n}_q$ can be deduced by the height ratio $R$ between the $\Delta n_q=-1$ and $\Delta n_q=1$ sidebands by $R=\bar{n}_q/(\bar{n}_q+1)$. Thus, we obtain the average phonon number on the three axes as $(\bar{n}_{x}, \bar{n}_{y},\bar{n}_{z})=(4.25\pm2.23, 2.61\pm1.01, 5.33\pm2.63)$. 
The corresponding temperatures are $(T_{x}, T_{y}, T_{z})=(15.8, 8.8, 9.0)$ $\mu$K. The temperature in the y direction and z direction is in good agreement with the temperature ($\sim 10$ $\mu$K) obtained by the method of release and recapture \cite{Release_and_recapture}. The temperature in the x direction is slightly higher because the optical pumping beam is aligned in this direction and heats the atom slightly in the state preparing process.

The effective Lamb-Dicke parameters are essential factors for assessing the effectiveness of an RSC. The effective Lamb-Dicke parameters can be obtained by 
\begin{equation}
\eta_{q,\text{eff}}=\eta_{q} \sqrt{2n_q+1}. 
\label{Lamb-Dicke}
\end{equation}
By using the initial phonon numbers obtained above, we obtain the effective Lamb-Dicke parameters for the Raman and optical pumping beams. They are $(\eta^{\mathrm{R}}_{x,\text{eff}}, \eta^{\mathrm{R}}_{y,\text{eff}}, \eta^{\mathrm{R}}_{z,\text{eff}}) \approx (0.52,0.62,0.78)$
and $(\eta^{\mathrm{OP}}_{x,\text{eff}}, \eta^{\mathrm{OP}}_{y,\text{eff}}, \eta^{\mathrm{OP}}_{z,\text{eff}})\approx (0.56, 0.46, 0.86)$. The initial effective Lamb-Dicke parameters are less than one, which indicates the effectiveness of the RSC. We will show this by measuring the average phonon number after applying the Raman cooling pulse sequence.

\begin{table}
\centering
\caption{\label{tab1} The dependence of the residual phonon numbers on the number of applied cooling cycles.}
\begin{tabular}{cccc}
\hline
Cooling cycles& $\bar{n}_x$ & $\bar{n}_y$ & $\bar{n}_z$\\
\hline
0 & $4.25^{+2.23}_{-2.23}$ & $2.61^{+1.01}_{-1.01}$ & $5.33^{+2.63}_{-2.63}$\\
10 & $0.31^{+0.19}_{-0.19}$ & $0.30^{+0.10}_{-0.10}$ & $0.37^{+0.16}_{-0.16}$\\
20 & $0.11^{+0.14}_{-0.11}$ & $0.19^{+0.10}_{-0.10}$ & $0.15^{+0.10}_{-0.10}$\\
30 & $0.02^{+0.10}_{-0.02}$ & $0.14^{+0.07}_{-0.07}$ & $0.12^{+0.09}_{-0.09}$\\
40 & $0.15^{+0.13}_{-0.13}$ & $0.12^{+0.08}_{-0.08}$ & $0.05^{+0.08}_{-0.05}$\\
50 & $0.07^{+0.07}_{-0.07}$ & $0.04^{+0.05}_{-0.04}$ & $0.08^{+0.06}_{-0.06}$\\
\hline

\end{tabular}
\end{table}

Table \ref{tab1} displays the residual phonon numbers on the three axes versus the number of Raman cooling cycles.The phonon numbers decrease very quickly for the first ten cooling cycles and turn slow for the following cooling cycles. The reason is the initial thermal distribution of the atom in the phonon states, which obeys a Bose-Einstein distribution with $P(n_q)=\bar{n}_q^{n_q}/(1+\bar{n}_q)^{n_q+1}$. The population in the ground state $|0_q\rangle$ is maximized, and it decreases with increasing phonon number $n_q$. 
The RSC pulse is most efficient for $|1_q\rangle$ state. The efficiency decreases with increasing $n_q$ due to the $n_q$-dependent Rabi frequency [Eq. (\ref{n=-1})] and the incremental effective Lamb-Dicke parameters [Eq. (\ref{Lamb-Dicke})]. Therefore, phonon states with smaller $n_q$ will rapidly cool to the $|0_q\rangle$ state during the initial cooling cycles. The cooling for phonon states with larger $n_q$ will be much slower.

Nevertheless, a single Cs atom can be efficiently prepared to its ground phonon state by 50 cooling cycles in 50 ms. The sideband spectra are displayed in Fig. \ref{spectroscopy}(c) and (d). The $n_q=-1$ sidebands are suppressed dramatically, which means that the atom mostly occupies the phonon ground state. The residual phonon numbers are $(\bar{n}_x,\bar{n}_{y},\bar{n}_z) \simeq (0.07^{+0.07}_{-0.07},0.04^{+0.05}_{-0.04},0.08^{+0.06}_{-0.06})$. The theoretical limits for the RSC are determined by the off-resonant Raman transitions and off-resonant spontaneous emission from the Raman beam \cite{PhysRevLett.75.4011}. They are $(\bar{n}_{x,lim},\bar{n}_{y,lim},\bar{n}_{z,lim}) \simeq (0.01,0.03,0.07)$ . The residual phonon numbers in the $y$ and $z$ directions approach the limit, but the residual phonon number in the $x$ direction is higher than the theoretical limit. The reason might be that the optical pumping beam is only in the $x$ direction and every spontaneous emission is accompanied by the absorption of one photon in this direction. Nevertheless, the final ground state populations in all three directions are $(P_{0x}, P_{0y}, P_{0z}) = (0.93,0.96,0.92)$, and the overall population in the 3D ground states is 82\%.

In conclusion, we have realized Raman sideband cooling of a single optically trapped Cs atom with an improved resolved RSC scheme. The scheme mitigates the influence of drift on the magnetic field and guarantees the long-term efficiency of the RSC. The 3D vibrational ground state was experimentally prepared with a probability of 82\%. This number is mainly restricted by the limited Lamb-Dicke parameters and can be increased by using a deeper trap with a smaller size and a lower initial temperature. The scheme can be used to cool other atoms with abundant Zeeman sublevels that are susceptible to magnetic fields.
 
\begin{backmatter}
\bmsection{Funding} This work was supported by the National Key Research and Development Program of China (Grant No. 2021YFA1402002), the National Natural Science Foundation of China (Grant Nos. U21A6006, U21A20433, 11974223, 11974225, 92265108, 12104277, and 12104278), the Fund for Shanxi 1331 Project Key Subjects Construction, and Fundamental Research Program of Shanxi Province (202203021223003).


\smallskip

\bmsection{Disclosures} The authors declare no conflicts of interest.

\bmsection{Data Availability Statement} No data were generated or analyzed in the presented research.

\end{backmatter}

\bibliography{sample}

\bibliographyfullrefs{sample}


\ifthenelse{\equal{\journalref}{aop}}{%
\section*{Author Biographies}
\begingroup
\setlength\intextsep{0pt}
\begin{minipage}[t][6.3cm][t]{1.0\textwidth} 
  \begin{wrapfigure}{L}{0.25\textwidth}
    \includegraphics[width=0.25\textwidth]{john_smith.eps}
  \end{wrapfigure}
  \noindent
  {\bfseries John Smith} received his BSc (Mathematics) in 2000 from The University of Maryland. His research interests include lasers and optics.
\end{minipage}
\begin{minipage}{1.0\textwidth}
  \begin{wrapfigure}{L}{0.25\textwidth}
    \includegraphics[width=0.25\textwidth]{alice_smith.eps}
  \end{wrapfigure}
  \noindent
  {\bfseries Alice Smith} also received her BSc (Mathematics) in 2000 from The University of Maryland. Her research interests also include lasers and optics.
\end{minipage}
\endgroup
}{}

\end{document}